\documentclass[superscriptaddress,preprint]{revtex4}
\usepackage{mathrsfs}
\usepackage{amssymb}
\usepackage[tbtags]{amsmath}
\usepackage{graphicx}
\usepackage{epsfig,graphicx,times}
\usepackage{color}

\begin{document}

\title{Jaynes-Cummings Models with trapped surface-state electrons in THz cavities}
\author{Miao Zhang, H.Y. Jia, J.S. Huang and L.F. Wei\footnote{weilianfu@gmail.com}}
\affiliation{Quantum Optoelectronics Laboratory, Southwest Jiaotong
University, Chengdu 610031, China}
\date{\today}

\begin{abstract}
An electron floating on the liquid Helium is proposed to be trapped
(by a micro-electrode set below the liquid Helium) in a high finesse
cavity.
Two lowest levels of the {\it vertical} motion of the electron acts
as a two-level ``atom", which could resonantly interact with the THz
cavity.
In the Lamb-Dicke regime, wherein the electron's in-plane activity
region is much smaller than the wavelength of the cavity mode, the
famous Jaynes-Cummings model (JCM) could be realized.
By applying an additional external classical laser beam to the
electron, a driven JCM could also be implemented. With such a driven
JCM certain quantum states, e.g., coherent states and the
Schr$\ddot{\text{o}}$dinger cat states, of the THz cavity field
could be prepared by one-step evolution.
The numerical results show that, for the typical parameters of the
cavity and electron on liquid Helium, a strong coupling between the
artificial atom and the THz cavity could be obtained.

PACS number(s): 42.50.Dv, 42.50.Ct, 73.20.-r.
\end{abstract}

\maketitle

\section{Introduction}
Jaynes-Cummings (JC) model (JCM), describing the basic interaction
of a two-level atom with a quantized electromagnetic field, is a
cornerstone to describe the interaction between light and matter in
{\it Quantum Optics}~\cite{Jaynes-Cummings}.
This famous model can explain many quantum phenomena and can be
implicated in recent quantum-state engineering and quantum
information processing (see,
e.g.,~\cite{JCM-Fock,JCM-entangled,JCM-gates}).
Usually, JCMs are implemented with the natural atoms in the cavity
QED systems (see, e.g.,~\cite{Rev.Mod.cavityQED}). There, the
cavities refer to the quantized optical or microwave resonators, and
the natural atoms are usually prepared at certain Rydberg
states~\cite{Science-cavity-2002,Nature-cavity-2003}. During the
flying atoms crossing the cavity, the JC interaction between two
selected internal electronic levels of the atoms and the cavity mode
can be realized.
Certainly, there has been also interested to realize the JC
Hamiltonian with other physical systems such as the trapped
ions~\cite{JCMs-trapped-ion}.

Electrons on the liquid Helium (called usually the surface-state
electrons) is one of the promise candidates, introduced first by
Platzman and Dykman~\cite{Science,PRB}, to implement quantum
information processing. Electrons on the surface of liquid Helium
are trapped in a set of one-dimensional ($1$D) hydrogenic levels by
their dielectric image potentials, and laterally confined by the
voltage on the micro-electrodes set below the liquid Helium.
A set of electrons trapped on the liquid Helium are effectively
coupled together via their Coulomb interactions. By applying
microwave radiation to these surface-state electrons from the
micro-electrodes, their ``atomic" states (acting as qubits) could be
coherently controlled.
Due to its scalability, easy manipulation, and relative long
coherence time, this system has been paid much attention in recent
years to implement quantum information processing (see,
e.g.,~\cite{Science,PRB,electron1,APL,electron2,Miaozhang-JCMs}).

In a recent work~\cite{Miaozhang-JCMs}, we have first shown that,
under the drivings of the applied laser beams, two directional
motions (i.e., the vertical direction's ``artificial" hydrogen and
the in-plane bosonic modes) of the trapped electrons could be
coupled together to generate the desirable JCMs. This is similar to
the laser-assisted coupling between the internal electronic and
external motional states of trapped ions~\cite{Rev.Mod.trapped-ion}.
In order to implement JCMs with significantly strong coupling in the
system of electrons floating on liquid Helium, here a THz cavity,
instead the quantized in-plane vibration of the electron, is
introduced to {\it resonantly} couple the ``artificial" hydrogen.
The configuration considered here is a single surface-state electron
trapped in a THz cavity. The cavity-induced interaction between the
electron's in-plane oscillation (the relevant vibrational region is
much smaller than the wavelength of the cavity mode) and the atomic
levels is negligible. Thus, the ``artificial" hydrogen could only be
resonantly coupled to the quantized THz mode and the desirable JCM
with strong coupling could be realized.
Furthermore, a driven JCM~\cite{Displaced fock state} could also be
generated by additionally applying an external classical laser beam
to drive the electron. With such a driven JCM, we show that some
nonclassical quantum states, such as the coherent states and
Schr$\ddot{\text{o}}$dinger cat states, of the THz cavity could be
easily prepared.

\section{JCM with a surface-state electron trapped in a THz cavity}
A single electron floating on the surface of liquid Helium (e.g.,
$^{4}\text{He}$) is weakly attracted by the dielectric image
potential $V(z)=-\Lambda e^2/z$, where $e$ is electron (with mass
$m_e$) charge, $z$ is the distance above liquid Helium surface, and
$\Lambda=(\varepsilon-1)/4(\varepsilon+1)=0.0069$ with
$\varepsilon=1.0568$ being dielectric constant of liquid
$^{4}\text{He}$~\cite{Rev.Mod.electrons}. A barrier (due to the
Pauli force) about $1$~eV is formed to prevent the electron
penetrates into the liquid Helium. Thus, the electron's motion
normal to the liquid Helium surface can be approximately described
by a $1$D hydrogen.
The energy levels associated with this motion form a hydrogen-like
spectrums $E_n=-\Lambda^2e^4m_e/2n^2\hbar^2$, which has been
experimentally observed~\cite{experiment-levels}. The corresponding
wave functions read~\cite{eigen-states}
\begin{equation}
\psi_n(z)=2n^{-\frac{5}{2}}r_B^{-\frac{3}{2}}
z\exp[-\frac{z}{nr_B}]L_{n-1}^{(1)}(\frac{2z}{nr_B}),
\end{equation}
with the Bohr radius $r_B=\hbar^2/(m_ee^2\Lambda)\approx76$~{\AA},
and Laguerre polynomials
\begin{equation}
L_n^{(\alpha)}(x)=\frac{e^xx^{-\alpha}}{n!}\frac{d^n}{dx^n}[e^{-x}x^{n+\alpha}].
\end{equation}

In the plane of the liquid Helium surface, the electron could be
confined by an potential generated by the charge $Q$ on the
micro-electrode, which is located at $h$ beneath the liquid Helium
surface (see Fig.1).
For simplicity, on the liquid Helium surface the electron is assumed
to be effectively constrained to move only along the $x$-axes.
Therefore, under the usual condition $z, x<< h$, the total potential
of the electron can be effectively approximated as~\cite{PRB}
\begin{equation}
U(z,x)\approx-\frac{\Lambda e^2}{z}+eE_\bot z+\frac{1}{2}m_e\nu^2x^2
\end{equation}
with $E_\perp\approx Q/h^2$ and
$\nu\approx\sqrt{eQ/(m_eh^3)}=\sqrt{eE_\perp/(m_eh)}$\,. This
indicates that the motions of the trapped electron can be regarded
as a 1D {\it Stark-shifted} hydrogen along the $z$-direction, and a
harmonic oscillation along the $x$-direction.
Following Dykman et.al.~\cite{PRB}, only two lowest levels (i.e.,
the ground state $|g\rangle$ and first excited state $|e\rangle$) of
the 1D hydrogen are considered to generate a qubit. As a
consequence, the Hamiltonian describing the electron reads
\begin{equation}
\hat{H}_e=\hbar\nu(\hat{a}^\dagger\hat{a}+\frac{1}{2})+\frac{\hbar
\omega_a}{2}\hat{\sigma}_z.
\end{equation}
Here, $\hat{a}^\dagger$ and $\hat{a}$ are the bosonic creation and
annihilation operators of the vibrational quanta (with frequency
$\nu$) of the electron's oscillation along the $x$-direction.
$\hat{\sigma}_z=|e\rangle\langle e|-|g\rangle\langle g|$ is the
Pauli operator. The transition frequency $\omega_a$ is defined by
$\omega_a=(E_e-E_g)/\hbar$ with $E_g$ and $E_e$ being the
corresponding energies of the lowest two levels, respectively.

\begin{figure}[tbp]
\includegraphics[width=9cm]{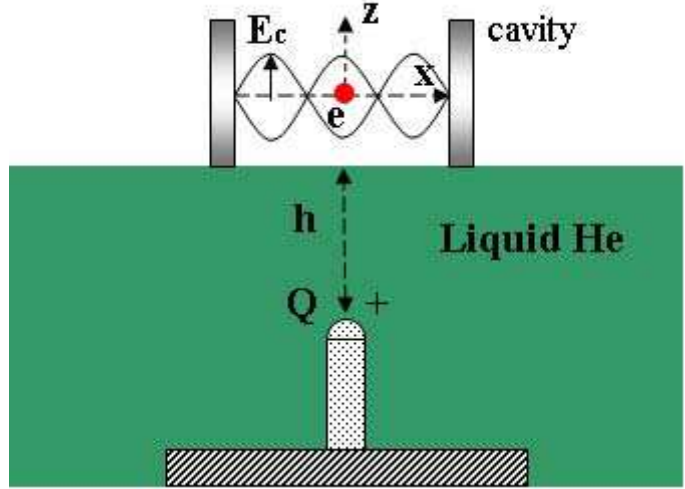}
\caption{(Color online) A sketch of a surface-state electron
confined in a high finesse cavity by a micro-electrode $Q$ submerged
by the depth $h$ beneath the Helium surface.}
\end{figure}

We now suppose that the above surface-state electron is trapped just
right in a QED cavity, see Fig.1. For simplicity, we assume that a
standing wave propagate on the $x$ axis and the electric vector of
the cavity takes the form~\cite{Quantum optics}
\begin{equation}
E_{c}(x,t)=\sqrt{\frac{\hbar\omega_c}{2\epsilon_0V}}e_z(\hat{b}^\dagger+\hat{b})\cos(k_cx+\phi_c).
\end{equation}
Here, $\epsilon_0$ is the vacuum dielectric constant, $\omega_c$ is
the frequency of the electric field, $V$ is the normalized volume of
the field, $e_z$ represents the unit polarized vector of
$z$-direction, $k_c$ is the wave number, $x$ denotes the position of
the oscillating electron from its trapped center, $\phi_c$ accounts
for the relative position of the center of the electron to the
standing wave and $\phi_c=0$ means that the ion is centered at an
antinode of the standing wave, $\hat{b}^\dagger$ and $\hat{b}$ are
the creation and annihilation operators of the basic mode of the
cavity field, respectively.

The Hamiltonian describing the whole motion of the electron now
reads
\begin{equation}
\hat{H}=\hat{H}_0+ezE_c(x,t),
\end{equation}
with
\begin{equation}
\hat{H}_0=\hbar\nu(\hat{a}^\dagger\hat{a}+\frac{1}{2})+\frac{\hbar
\omega_a}{2}\hat{\sigma}_z+\hbar\omega_c\hat{b}^\dagger\hat{b}.
\end{equation}
Certainly, a more realistic model should include the cavity mode and
electron's interaction with a dissipative environment (e.g., the
liquid Helium and the finite quality factor cavity). Here, we are
interested in the strong coupling regime wherein the dissipation
could be neglected. With the Pauli and bosonic operators defined
above, the positions $z$ and $x$ can be expressed as
\begin{equation}
\hat{z}=z_{ge}\hat{\sigma}_x+\frac{1}{2}(z_{ee}-z_{gg})\hat{\sigma}_z
\,\,\,\,\text{and}\,\,\,\,\,
x=\sqrt{\frac{\hbar}{2m_e\nu}}(\hat{a}^\dagger+\hat{a}),
\end{equation}
where $z_{ee}=\langle e|z|e\rangle$, $z_{gg}=\langle g|z|g\rangle$,
and $z_{ge}=\langle e|z|g\rangle$. As a consequence, the Hamiltonian
(6) can be rewritten as ($\phi_c=0$)
\begin{equation}
\hat{H}=\hat{H}_0
+\hbar\Omega_c(\hat{b}^\dagger+\hat{b})\hat{\sigma}_x
(e^{i\eta_c(\hat{a}^\dagger+\hat{a})}+e^{-i\eta_c(\hat{a}^\dagger+\hat{a})})
+\hbar\widetilde{\Omega}_c(\hat{b}^\dagger+\hat{b})\hat{\sigma}_z
(e^{i\eta_c(\hat{a}^\dagger+\hat{a})}+e^{-i\eta_c(\hat{a}^\dagger+\hat{a})})
\end{equation}
where $\Omega_c=ez_{ge}\sqrt{\omega_c/8\hbar\epsilon_0V}$ describes
the strength of coupling between the electron and cavity field.
$\widetilde{\Omega}_{c}=e(z_{ee}-z_{gg})\sqrt{\omega_c/32\hbar\epsilon_0V}\neq0$,
due to {\it the broken parities} of the quantum states of the above
1D hydrogen. Finally, $\eta_c=\omega_c\sqrt{\hbar/(2m_e\nu)}/c$
(where $c$ is the velocity of light) is the so-called LD parameter
describing the strength of the cavity field induced coupling between
the $z$- and $x$-direction motions of the trapped electron.

For the typical parameters $E_\perp=3\times10^{4}$~V/m and
$h=5\times10^{-7}$~m~\cite{PRB}, the $z$-direction transition
frequency and the $x$-direction's vibrational frequency of the
electron are estimated as $\omega_a/(2\pi)\approx0.27$~THz and
$\nu/(2\pi)\approx16$~GHz, respectively. Consequently, the above LD
parameter $\eta_c\approx10^{-4}$ for the resonant condition:
$\omega_c=\omega_a$. For such a sufficiently small LD parameter, one
can perform an approximation $\exp[\pm
i\eta(\hat{a}+\hat{a}^\dagger)]\approx1$ (which means that the
cavity-induced interaction between the $z$- and $x$-direction's
motions of the trapped electron is robustly neglected). Therefore,
the Hamiltonian (9) can be further simplified to
\begin{equation}
\hat{H}_L=\hat{H}_0
+2\hbar\Omega_c(\hat{b}^\dagger+\hat{b})\hat{\sigma}_x
+2\hbar\widetilde{\Omega}_c(\hat{b}^\dagger+\hat{b})\hat{\sigma}_z.
\end{equation}
In the interaction picture defined by
$\hat{U}=\exp(-it\hat{H}_0/\hbar)$, we have
\begin{equation}
\hat{H}_I=2\hbar\Omega_c(\hat{b}^\dagger
e^{i\omega_ct}+\hat{b}e^{-i\omega_ct})
(\hat{\sigma}_-e^{-i\omega_at}+\hat{\sigma}_+e^{i\omega_at})
+2\hbar\widetilde{\Omega}_c\hat{\sigma}_z(\hat{b}^\dagger
e^{i\omega_ct}+\hat{b}e^{-i\omega_ct}).
\end{equation}
Assume that the applied the cavity is resonant with the atomic
qubit, i.e., $\omega_a=\omega_c$, the famous JCM
\begin{equation}
\hat{H}_{\text{JC}}=2\hbar\Omega_c(\hat{b}^\dagger\hat{\sigma}_-
+\hat{b}\hat{\sigma}_+)
\end{equation}
could be obtained under the usual rotating-wave approximation
(RWA)~\cite{RWA,Weigates}. For an initial state $|m\rangle|g\rangle$
or $|m\rangle|e\rangle$ the time-evolution under the Hamiltonian
(12) reads
\begin{eqnarray}
\left\{
\begin{array}{l}
|m\rangle|g\rangle\longrightarrow\cos(\Omega_{m-1}t)|m\rangle|g\rangle
-i\sin(\Omega_{m-1}t)|m-1\rangle|e\rangle,\\
|m\rangle|e\rangle\longrightarrow\cos(\Omega_{m}t)|m\rangle|e\rangle
-i\sin(\Omega_{m}t)|m+1\rangle|g\rangle,
\end{array}
\right.
\end{eqnarray}
with
\begin{eqnarray}
\Omega_{m}=\left\{
\begin{array}{l}
2\Omega_c\sqrt{m+1}\,\,\,\,\,\,\text{for}\,\,\,\,\,\,m\geq0,\\
0\,\,\,\,\,\,\,\,\,\,\,\,\,\,\,\,\,\,\,\,\,\,\,\,\,\,\,\,\,\,\,\,\,
\text{for}\,\,\,\,\,\,m<0
\end{array}
\right.
\end{eqnarray}
being the effective Rabi frequency, which depends obviously on the
initial occupation number $m$ of the cavity. With such a JCM, the
coherent vacuum Rabi oscillation~\cite{Rev.Mod.cavityQED} between
the present artificial qubit and the vacuum cavity could be
presented.


\section{A driven JCM implemented by additionally applying a classical laser beam}

Now, we apply an additional classical laser beam:
$E_l(x,t)=E_ze_z\cos(k_lx-\omega_lt-\phi_l)$ to the floating
electron, and then write the total Hamiltonian describing the
present model as
\begin{equation}
\begin{array}{l}
\hat{H}=\hat{H}_0+eE_c(x,t)z+eE_l(x,t)z
\\
\,\,\,\,\,\,\,
=\hat{H}_0+\hbar\Omega_c(\hat{b}^\dagger+\hat{b})\hat{\sigma}_x
(e^{i\eta_c(\hat{a}^\dagger+\hat{a})}+e^{-i\eta_c(\hat{a}^\dagger+\hat{a})})
\\
\,\,\,\,\,\,\,\,\,\,\,\,\,+
\hbar\widetilde{\Omega}_c(\hat{b}^\dagger+\hat{b})\hat{\sigma}_z
(e^{i\eta_c(\hat{a}^\dagger+\hat{a})}+e^{-i\eta_c(\hat{a}^\dagger+\hat{a})})

\\
\,\,\,\,\,\,\,\,\,\,\,\,\,+\hbar\Omega_l\hat{\sigma}_x
(e^{i\eta_l(\hat{a}+\hat{a}^\dagger)-i\omega_lt-i\phi_l}
+e^{-i\eta_l(\hat{a}+\hat{a}^\dagger)+i\omega_lt+i\phi_l})
\\
\,\,\,\,\,\,\,\,\,\,\,\,\,+
\hbar\widetilde{\Omega}_l\hat{\sigma}_z(e^{i\eta_l(\hat{a}+\hat{a}^\dagger)-i\omega_lt-i\phi_l}
+e^{-i\eta_l(\hat{a}+\hat{a}^\dagger)+i\omega_lt+i\phi_l}).
\end{array}
\end{equation}
Here, $E_z$, $e_z$, $k_l$, $\omega_l$, and $\phi_l$ are the
amplitude, unit polarized vector of $z$-direction, wave-vector,
frequency, and initial phase of the applied classical laser beam,
respectively. $\Omega_l=ez_{ge}E_z/(2\hbar)$ describes the coupling
strength between the electron and the applied classical laser field.
$\widetilde{\Omega}_l=e(z_{ee}-z_{gg})E_z/(4\hbar)\neq0$ due to
again the broken parities of the quantum states of the 1D hydrogen.
Also, the new LD parameter $\eta_l=\omega_l\sqrt{\hbar/(2m_e\nu)}/c$
describes the strength of the laser induced coupling between the
motions of $z$-and $x$-directions of the trapped electron.

Again, under the resonant condition: $\omega_c=\omega_l=\omega_a$,
all the relevant LD parameters are significantly small, i.e.,
$\eta_c=\eta_l\approx10^{-4}\sim 0$. As a consequence, $\exp[\pm
i\eta(\hat{a}+\hat{a}^\dagger)]\approx 1$, and thus the Hamiltonian
(15) can be simplified to
\begin{equation}
\begin{array}{l}
\hat{H}_L=\hat{H}_0
+2\hbar\Omega_c(\hat{b}^\dagger+\hat{b})\hat{\sigma}_x
+2\hbar\widetilde{\Omega}_c(\hat{b}^\dagger+\hat{b})\hat{\sigma}_z
\\
\,\,\,\,\,\,\,\,\,\,\,\,\,+\hbar\Omega_l\hat{\sigma}_x
(e^{-i\omega_lt-i\phi_l}+e^{i\omega_lt+i\phi_l})
+\hbar\widetilde{\Omega}_l\hat{\sigma}_z
(e^{-i\omega_lt-i\phi_l}+e^{i\omega_lt+i\phi_l}),
\end{array}
\end{equation}
which reduces to
\begin{equation}
\begin{array}{l}
\hat{H}_I=2\hbar\Omega_c(\hat{b}^\dagger
e^{i\omega_ct}+\hat{b}e^{-i\omega_ct})
(\hat{\sigma}_-e^{-i\omega_at}+\hat{\sigma}_+e^{i\omega_at})
+2\hbar\widetilde{\Omega}_c\hat{\sigma}_z(\hat{b}^\dagger
e^{i\omega_ct}+\hat{b}e^{-i\omega_ct})
\\
\,\,\,\,\,\,\,\,\,\,\,\,\,+\hbar\Omega_l\hat{\sigma}_-
(e^{-i(\omega_l+\omega_a)t-i\phi_l}+e^{i(\omega_l-\omega_a)t+i\phi_l})
+\hbar\Omega_l\hat{\sigma}_+
(e^{i(\omega_a-\omega_l)t-i\phi_l}+e^{i(\omega_l+\omega_a)t+i\phi_l})
\\
\,\,\,\,\,\,\,\,\,\,\,\,\, + \hbar\widetilde{\Omega}_l\hat{\sigma}_z
(e^{-i\omega_lt-i\phi_l}+e^{i\omega_lt+i\phi_l}).
\end{array}
\end{equation}
in the interaction picture defined by
$\hat{U}=\exp(-it\hat{H}_0/\hbar)$.

Under the above resonant condition and the usual RWA, a driven
JCM~\cite{Displaced fock state}:
\begin{equation}
\hat{H}_{\text{DJC}}=2\hbar\Omega_c(\hat{b}^\dagger\hat{\sigma}_-
+\hat{b}\hat{\sigma}_+)+\hbar\Omega_{l}(e^{i\phi_l}\hat{\sigma}_-
+e^{-i\phi_l}\hat{\sigma}_+),
\end{equation}
is realized. Where, the first term corresponds to the JC
interaction, and the second term is due to the driving of the
classical laser field. The above driven JCM is solvable and the
relevant time-evolution operator reads
\begin{equation}
\hat{U}(t)=e^{-\frac{it}{\hbar}D^\dagger(r)\hat{H}_{\text{JC}}D(r)}
=D(-r)e^{-\frac{it}{\hbar}\hat{H}_{\text{JC}}}D(r).
\end{equation}
Here, $D(r)=\exp(r\hat{b}^\dagger-r^*\hat{b})$ is the displacement
operator with $r=e^{-i\phi_l}\Omega_{l}/(2\Omega_c)$. One can easily
prove that
\begin{equation}
D^\dagger(r)=D^{-1}(r)=D(-r),\,\,\,\,\,
D^\dagger(r)\hat{b}D(r)=\hat{b}+r,\,\,\,\,\,D^\dagger(r)\hat{b}^\dagger
D(r)=\hat{b}^\dagger+r^*.
\end{equation}

A potential application of the driven JCM proposed here is that it
can be utilized to prepare certain typical nonclassical quantum
states (e.g., the displaced Fock states~\cite{Displaced fock state}
and Schr$\ddot{\text{o}}$dinger cat states~\cite{cat state}) of the
THz cavity. In order to implement these preparations, we first set
$\phi_l=0$ and perform a basis transformation
$|\pm\rangle=(|g\rangle\pm|e\rangle)/\sqrt{2}$ to rewrite the driven
JCM Hamiltonian as
\begin{equation}
\hat{H}'_{\text{DJC}}=\hbar\Omega_c\left[\hat{b}^\dagger(\hat{\tau}_z-\hat{\tau}_++\hat{\tau}_-)
+\hat{b}(\hat{\tau}_z+\hat{\tau}_+-\hat{\tau}_-)\right]+\hbar\Omega_{l}\hat{\tau}_z
\end{equation}
with $\hat{\tau}_z=\hat{\sigma}_x$,
$\hat{\tau}_+=(-\hat{\sigma}_z-\hat{\sigma}_-+\hat{\sigma}_+)/2$ and
$\hat{\tau}_-=(-\hat{\sigma}_z+\hat{\sigma}_--\hat{\sigma}_+)/2$. In
the interaction picture defined by
$\hat{U}=\exp(-i\Omega_lt\hat{\tau}_z)$, we have
\begin{equation}
\hat{H}''_{\text{DJC}}=\hbar\Omega_c
\left[\hat{b}^\dagger(\hat{\tau}_z-e^{i2\Omega_lt}\hat{\tau}_++e^{-i2\Omega_lt}\hat{\tau}_-)
+\hat{b}(\hat{\tau}_z+e^{i2\Omega_lt}\hat{\tau}_+-e^{-i2\Omega_lt}\hat{\tau}_-)\right].
\end{equation}
Note that the amplitude $E_z$ of the applied laser field is
experimentally controllable, and thus $\Omega_l$ can be effectively
increased with the increasing of $E_z$. Next, let us consider the
strong driving case where $\Omega_l\gg\Omega_c$. Under the RWA, the
Hamiltonian (22) yields the combination of a JC and an anti-JC
interactions~\cite{cat state}
\begin{equation}
\hat{H}_{\text{eff}}=\hbar\Omega_c(\hat{b}^\dagger
+\hat{b})\hat{\tau}_z=\hbar\Omega_c(\hat{b}^\dagger\hat{\sigma}_-
+\hat{b}\hat{\sigma}_+)+\hbar\Omega_c(\hat{b}^\dagger\hat{\sigma}_+
+\hat{b}\hat{\sigma}_-).
\end{equation}
Obviously, if the electron-cavity is initially prepared in
$|0\rangle|+\rangle$, then a coherent state
$|\alpha\rangle=\exp(-|\alpha|^2/2)\sum_{m=0}^\infty(\alpha^m/\sqrt{m!})|m\rangle,\,\alpha=-it\Omega_c$
of the THz cavity field can be generated.
On the other hand, if the electron-cavity is initially prepared in
$|0\rangle|g\rangle=|0\rangle(|+\rangle+|-\rangle)/\sqrt{2}$, then a
Schr$\ddot{\text{o}}$dinger cat state of the THz cavity:
\begin{equation}
|\varphi\rangle=\frac{1}{\sqrt{2}}\left(|\alpha\rangle|+\rangle+|-\alpha\rangle|-\rangle\right)
=\frac{1}{2}
\left[(|\alpha\rangle+|-\alpha\rangle)|g\rangle+(|\alpha\rangle-|-\alpha\rangle)|e\rangle\right]
\end{equation}
is implemented. Furthermore, if the electron is detected in
$|g\rangle$, then the cavity field collapses to the so-called even
coherent states; whereas if the electron is detected in $|e\rangle$,
then the cavity field collapses to the so-called odd coherent
states.

\section{Discussions}
In principle, the device proposed here is experimental feasibility.
First, trapping electrons in a small region on the surface of liquid
Helium by setting suitable micro-electrodes has been experimentally
realized~\cite{APL}. For the typical vibrational frequency
$\nu/(2\pi)\approx16$~GHz and transition frequency
$\omega_a/(2\pi)\approx0.27$~THz, the localization length of the
electron moving in the plane is
$L_\parallel=\sqrt{\hbar/m_e\nu}\approx0.3$~nm, which is far less
than that of the size of the corresponding cavity (it is on the
order of wavelength $2\pi c/\omega_a\approx1$~mm). Also, the
vertical motions of the electron (which is estimated on the order of
$r_B\approx76$~{\AA}) is far less the typical value of cavity waist
(probably on the order of $\mu$m).

Second, the superfluid Helium ($<2.2$K) naturally provides a
sufficiently low temperature surrounding of the present system. As
consequence, the thermal noise of the cavity could be well
suppressed. For example, for the thermal state
\begin{equation}
\rho=\sum_{m=0}^\infty
[1-e^{-\hbar\omega_c/(k_BT)}]e^{-m\hbar\omega_c/(k_BT)}|m\rangle\langle
m|
\end{equation}
most of photons are at the vacuum state, e.g., the probability $P_0$
of the photons at the vacuum state is significantly large:
$P_0>96\%$, for the cavity with frequency (e.g., $\sim 1$THz).
Above, $k_B$ and $T$ are the Boltzmann constant and temperature of
the cavity, respectively.

Thirdly, the system proposed here could work in a strong-coupling
regime, i,e., the interaction strength is sufficiently larger than
the decays of the artificial atom and the cavity.
In fact, the decay rate of the above artificial 1D hydrogen is
estimated~\cite{Science,PRB} to be $\gamma\sim 10$~KHz for the
typical vibrational frequencies $\nu/(2\pi)\sim 10$~GHz and
transition frequency $\omega_a/(2\pi)\sim 0.1$~THz. The main source
of the noises on the atom is the so-called ripplons, i.e, the
thermally excited surface waves of liquid Helium~\cite{PRB,Science}.
Probably, decay rate of the atom could be significantly decreased by
localizing electrons more strongly in the plane (corresponding to a
larger frequency of in-plane vibrations). Worthy of note, localizing
electrons more strongly in the plane of the liquid Helium surface
deceases also the LD parameters, making the LD approximations
performed above work better.
On the other hand, the decay of the cavity is mainly due to the
photon scattering or absorbing on the imperfect cavity mirrors. A
high finesse cavity can significantly decrease the decay rate, as
$\kappa\approx c\pi/(2FL)$. Here, $L$ and $F$ are the length and
finesse of the cavity, respectively. For example, the decay rate of
the Fabry-Perot cavity (with $L=0.12$~mm and $F=4.4\times10^5$) in
Ref.~\cite{FP-cavity} is as low as $\kappa=8.9$~MHz.
Slightly differing from the cavity in Ref.~\cite{FP-cavity}, let us
consider a cavity with a modest size, e.g., $L\approx1$~mm, which is
comparable to the typical wave length: $\lambda=2\pi c/\omega_a$, of
the present system $\omega_a/(2\pi)\approx0.27$~THz. The decay rate
of this cavity could be further lowered to be $\kappa=1.1$~MHz.
Also, the transition matrix element is estimated as $z_{ge}\approx
0.5r_B$ and the volume mode of the cavity is
$V=\pi(w/2)^2L\approx3.14\times10^{-4}\,\text{mm}^3$ for a typical
waist $w=20\,\mu$m~\cite{cavity-waist-numbers}.
Therefore, the so-called coupling strength could be up to
$g_0=2\Omega_c\approx 33\,\text{MHz}$, which is significantly larger
than the decay rates of the above atom and cavity. As a
consequence~\cite{cavity-waist-numbers}, the number of photons
inside the cavity needed to appreciably affect the electron is
calculated as $n_0=\gamma^2/2g_0^2\approx4.6\times10^{-8}$, and the
number of electrons needed to appreciably affect the cavity field is
$N_0=2\kappa\gamma/g_0^2\approx2\times10^{-5}$. This implies that
the present system could work in the strong-coupling regime.

Compared to the usual system of a natural atom interacting with an
microwave cavity~\cite{Rev.Mod.cavityQED}, one of the most
advantages of the present system is that it possesses a sufficiently
long electron-cavity interaction time. Indeed, the interaction time
between the flying atoms and microwave cavity are relatively short
(i.e., on the order of $\sim\mu s$~\cite{Rev.Mod.cavityQED}). In
order to increasing cavity-atom interaction time, the optical
lattice technique~(see, e.g.,~\cite{cavity-trap}) have been utilized
to trap the atoms in cavities, and the recent experiments showed
that such a time can be increased to the order of
$\text{s}$~\cite{cavity-trap-s}.
Like the trapped ions inside the cavity~\cite{trap-ions-cavity}, the
present system has sufficiently long electron-cavity interaction
time, because the surface-state electron can be, in principle,
always trapped in the cavity waist by the applied micro-electrode.

{\bf Acknowledgements}: This work is partly supported by the NSFC
grant No. 10874142, 90921010, and the grant from the Major State
Basic Research Development Program of China (973 Program) (No.
2010CB923104).



\end{document}